\documentclass[epj]{svjour}
\usepackage{textcomp}
\usepackage{amsmath}
\usepackage{amsfonts}
\usepackage{calc}
\usepackage{mathrsfs}
\usepackage{amssymb}
\usepackage{amsmath}
\usepackage{color}
\usepackage{bm}
\usepackage{graphicx}

\renewcommand{\k}{{\bm{k}}}
\newcommand{\m}{m^*_{\rm op}}
\newcommand{\n}{n_{\rm eff}}

\begin{document}
\title{Optical properties of the pseudogap state in underdoped cuprates} 
\author{Adam Pound\inst{1} \and J.P. Carbotte\inst{2,3} \and E.J. Nicol\inst{1}}
\institute{Department of Physics, University of Guelph, Guelph, Ontario, Canada, N1G 2W1\and
Department of Physics and Astronomy, McMaster University, Hamilton, Ontario, Canada, L8S 4M1\and
The Canadian Institute for Advanced Research, Toronto, Ontario, Canada, M5G 1Z8}
\date{\today}

\abstract{
Recent optical measurements of deeply underdoped cuprates have
revealed that a coherent Drude response persists well below the
end of the superconducting dome. In addition, no large increase in optical
effective mass has been observed, even at dopings as low as 1\%.
We show that this behavior is consistent with the resonating valence
bond spin-liquid model proposed by Yang, Rice, and Zhang.
In this model, the overall reduction in optical conductivity in the approach to the
Mott insulating state is caused not by an increase in effective mass, but by
a Gutzwiller factor, which describes decreased coherence due to correlations,
and by a shrinking of the Fermi surface, which decreases the number of available charge carriers.
We also show that in this model, the pseudogap does not modify the low-temperature,
low-frequency behavior, though the magnitude of the conductivity is greatly reduced by the Gutzwiller factor.
Similarly, the profile of the temperature dependence of the microwave conductivity
is largely unchanged in shape, but the Gutzwiller factor is essential in understanding
the observed difference in magnitude between ortho-I and -II YBa$_2$Cu$_3$O$_y$.
}

\maketitle

\section{Introduction}
As formulated by Yang, Rice, and Zhang (YRZ) \cite{yrz:2006}, the resonating valence bond model \cite{anderson:2007} of the superconducting underdoped cuprates has been remarkably successful in providing an understanding of properties previously considered anomalous \cite{valenzuela:2007,leblanc:2010,carbotte:2010,illes:2009,yrz:2009,yang:2010,borne:2010}. An essential element of the model is the formation of a pseudogap that grows in magnitude with decreased doping and reconstructs the Fermi surface into Luttinger hole and electron pockets, which replace the usual large Fermi surface of Fermi liquid theory. This extension of BCS theory is quite different from the many others that were required to explain the properties of conventional and even optimally and overdoped cuprates. Those extensions include anisotropy \cite{leung:CJP:1976}, energy dependence in the electronic density of states \cite{schachinger:1990}, inelastic scattering \cite{nicol-carbotte-timusk:1991,marsiglio:1996}, strong coupling effects \cite{schachinger:1980}, and $d$-wave gap symmetry with possibly higher harmonics \cite{o'donovan:1995a,o'donovan:1995b}.

In this paper, we use the YRZ model to explain anomalous optical conductivity data. W.J. Padilla et al. \cite{padilla:2005} and Y.S. Lee et al. \cite{lee:2005} recently measured the optical response in the deeply underdoped region of the high-$T_c$ cuprate phase diagram. In both La$_{2-x}$Sr$_x$CuO$_4$ (LSCO) and YBa$_2$\-Cu$_3$O$_y$ (YBCO), they found that a coherent Drude response, characteristic of metallicity, persisted well below the end of the superconducting dome in the approach to the antiferromagnetic Mott instulating state, even for doping values as low as $x=1\%$ in LSCO and $y=6.3$ in YBCO. They also noted that, remarkably, the optical effective mass of the charge carriers remains nearly constant, reaching at most a few times the bare electron mass. Hence, they concluded that the approach to the insulating state must be due to a reduction of the number of charge carriers rather than a decrease in carrier mobility. They suggested exotic theoretical models to explain this behavior; however, we show that their observations are fully consistent with the simpler YRZ model.

In Section~\ref{formalism} we summarize the basic elements of the model that are needed to calculate the AC optical conductivity. In Section~\ref{AC conductivity} we present formulas for AC conductivity at any temperature in the pseudogap state, and we present numerical results for a number of dopings. We focus on the  approach to the Mott insulating state, characterizing the elements that control the size of the remaining Drude response: the reduction in number of charge carriers, the reduction of the coherent response due to correlations, and the reduction in carrier mobility. In Section~\ref{DC conductivity} we consider optical properties in the superconducting state. We examine the effect of pseudogap formation on the thermal conductivity and low-frequency optical conductivity, showing modifications to the Wiedemann-Franz law at significant temperatures but no effect in the approach to the universal zero-temperature limits. We then consider the microwave conductivity, comparing numerical results to experimental data on ortho-I and -II YBCO; we find that the YRZ model provides qualitative agreement with the data, while Fermi liquid theory does not. In Section~\ref{summary} we summarize our findings.

\section{Formalism}\label{formalism}
In the phenomenological resonating valence bond spin liquid model of Yang, Rice, and Zhang, the coherent part of the charge carrier's Green's function is written as
\begin{equation}\label{Greens function}
G(\k,\omega) = \sum_{\alpha=\pm}\frac{g_t(x)W_\k^\alpha}{\omega-E^\alpha_\k-\Delta^2_{sc}(\k)/(\omega+E_\k^\alpha)},
\end{equation}
where the two energy branches are given by
\begin{align}
E^\pm_\k &= \frac{1}{2}(\xi_\k-\xi^0_\k)\pm E_\k,
\end{align}
with
\begin{align}
E_\k &= \sqrt{\tilde\xi_\k^2+\Delta^2_{pg}},\\
\tilde\xi_\k &= \frac{1}{2}(\xi_\k+\xi^0_\k).
\end{align}
The two branches have weights
\begin{align}
W^\pm_\k &= \frac{1}{2}\left(1\pm\frac{\tilde\xi_\k}{E_\k}\right).\label{Wpm}
\end{align}
In these formulas, $\k$ is momentum and $\omega$ is energy; $\Delta_{pg}$ is the pseudogap;
\begin{align}
\xi_\k&=-2t(x)(\cos k_xa+\cos k_ya)-4t'(x)\cos k_xa\cos k_ya\nonumber\\
&\quad-2t''(x)(\cos2k_xa+\cos2k_ya)-\mu_p
\end{align}
is the band-theory dispersion curve for a tight-binding Hamiltonian including contributions up to third-nearest-neighbor hopping, with a chemical potential $\mu_p$; and $\xi^0_\k=-2t(x)(\cos k_xa+\cos k_ya)$ is the first-nearest-neighbor contribution alone. The hopping parameters are $t(x) = g_t(x)t_0$ $+ \frac{3}{8}g_s(x)J\chi$, $t'(x) = g_t(x)t'_0$, and $t''(x) = g_t(x)t''_0$, where
\begin{equation}
g_t(x)=\frac{2x}{(1 + x)}\ \ {\rm and}\ \ g_s(x)=\frac{4}{(1 + x)^2}
\end{equation}
are Gutzwiller factors accounting for the loss of coherence due to correlations and narrowing of bands. The parameters $J$ and $\chi$ take values $J = \frac{1}{3}t_0$ and $\chi = 0.338$, and in accordance with YRZ, we use $t'_0=-0.3t_0$, and $t''_0 = 0.2t_0$, chosen to duplicate the energy dispersion of Ca$_2$CuO$_2$Cl$_2$. Both the pseudogap and the superconducting gap are taken to have $d$-wave symmetry, 
\begin{align}
\Delta_{pg} &=\frac{1}{2}\Delta^0_{pg}(x)(\cos k_xa-\cos k_ya),\\
\Delta_{sc} &= \frac{1}{2}\Delta^0_{sc}(x)(\cos k_xa-\cos k_ya),
\end{align}
with doping-dependent magnitudes
\begin{align} 
\Delta^0_{pg}(x) &= 0.6t_0(1-x/0.2),\\
\Delta^0_{sc}(x) &= 0.14t_0[1-82.6(x-0.2)^2].
\end{align}
These dependencies place both optimal doping and the quantum critical point for pseudogap formation at $x=0.2$. One could easily alter this placement, but our aim is to analyze qualitative features rather than fit the model to experiment. Superconductivity has been included at the level of BCS theory modified to account for $d$-wave gap symmetry, and we give the gap a temperature dependence to accord with that. We take the pseudogap to be independent of temperature.

One can use the Green's function \eqref{Greens function} to find the associated regular and anomalous spectral functions
\begin{align}
A(\k,\omega) &= \sum_{\alpha=\pm}g_tW^\alpha_\k\big[(u^\alpha)^2\delta(\omega-E^\alpha_S)\nonumber\\
&\quad+(v^\alpha)^2\delta(\omega+E^\alpha_S)\big],\\
B(\k,\omega) &= \sum_{\alpha=\pm}g_tW^\alpha_\k \frac{\Delta_{sc}}{2E_S^\alpha}\left[\delta(\omega-E^\alpha_S)-\delta(\omega+E^\alpha_S)\right]\!,\!
\end{align}
where
\begin{align}
u^\alpha &= \left[\frac{1}{2}\left(\!1+\frac{E^\alpha_\k}{E^\alpha_S}\right)\right]^{1/2},\\
v^\alpha &= \left[\frac{1}{2}\left(\!1-\frac{E^\alpha_\k}{E^\alpha_S}\right)\right]^{1/2}.
\end{align}
Here we see that there are four energy branches, $\pm E_S^\alpha$, where $E_S^\alpha=\sqrt{(E_\k^\alpha)+\Delta_{sc}^2}$. To account for impurity scattering, in all our calculations we broaden the Dirac delta functions in the spectral functions into a Lorentzian form with half-width $\Gamma/2$.

Many of our results in this paper can be understood from the manner in which the pseudogap in this model causes the large Fermi surface of Fermi liquid theory to reconstruct into a hole Luttinger pocket centered on the nodal direction and an electron pocket centered on the antinodal direction in the Brillouin zone. Each of these pockets has a strongly weighted side, which follows fairly closely the underlying Fermi liquid Fermi surface, and a weakly weighted side, which is more closely related to the antiferromagnetic Brillouin zone boundary defined by $\xi_\k^0=0$. As the doping is decreased, these pockets shrink; for sufficiently large values of the pseudogap (specifically, for values of $x\lesssim0.16$), the electron pockets are gapped out entirely. This situation is shown for the first quadrant of the Brillouin zone in the insets of the left-hand column of Fig.~\ref{basov}: at $x=0.2$, there is no pseudogap, and the Fermi surface is the large open contour of a Fermi liquid; at smaller values, the Fermi surface is reconstructed into ever-smaller pockets. But though these pockets shrink, they are present for all nonzero values of $x$. And on parts of the contours around these pockets, charge carriers maintain high mobility and low-energy excitations can easily occur. Hence, the hole contours always provide a source of mobile charge carriers and maintain coherence and metallicity in the system as the Mott insulating state is approached. As we shall see in the next section, this leads to a persistent Drude response in the AC optical conductivity. In addition, the reconstruction of the Fermi surface leaves the area around the node unaffected, preserving low-temperature, low-energy physics that accesses states only near the node. In Sec.~\ref{DC conductivity} we show that this leads to traditional zero-temperature universal limits persisting in the pseudogap state.

For our purposes, the Green's function in Eq.~\eqref{Greens function} can be viewed as phenomenogical, with its primary use being a qualitative understanding of the anomalous properties associated with pseudogap phenomena in underdoped cuprates. However, its form is based on a microscopic theory of arrays of two-legged Hubbard ladders in a doped spin liquid including long-range interladder hopping~\cite{yrz:2006}. And although it was initially considered only for weak coupling, comparison with the numerical results of Troyer et al.~\cite{troyer:1996} for strong coupling $t$-$J$ ladders suggests that it is characteristic of all doped spin liquids. Therefore, the physics encapsulated in Eq.~\eqref{Greens function} should apply to any materials that can be described by the microscopic Hubbard model near half-filling.

\section{AC conductivity and optical effective mass}\label{AC conductivity}
In this section we examine the optical conductivity as the insulating state is approached. Since that approach lies outside the superconducting dome, we focus on the non-superconducting state, even when analyzing dopings and temperatures that would fall within the dome in experiment. Now, the real part of the AC conductivity $\sigma=\sigma_1+i\sigma_2$ at frequency $\omega$ and temperature $T$ is given by
\begin{align}\label{sigma_1}
\sigma_1(\omega,T) &= -\frac{2\pi e^2}{\omega}\sum_\k v^2_{k_x}\int_{-\infty}^\infty d\omega'\left[f(\omega'+\omega)-f(\omega')\right]\nonumber\\
&\quad\times\Big[A(\k,\omega')A(\k,\omega'+\omega)\nonumber\\
&\quad+B(\k,\omega')B(\k,\omega'+\omega)\Big],
\end{align}
where $e$ is the magnitude of the electron's charge, $v_{k_x}$ is the $x$ component of the velocity in a state of momentum $\k$, and $f(\omega)$ is the Fermi function $1/[1+\exp(\omega/k_BT)]$. The sum over $\k$ extends over the first CuO$_2$ Brillouin zone. In this expression, we have neglected vertex corrections: the only effect of these is to change the overall magnitude of the conductivity, which is not essential to this work.

In the non-superconducting state, the imaginary part of $\sigma$ can be obtained from the real part via a Kramers-Kronig transformation. In the superconducting state, the formation of the superfluid transfers part of the conductivity into a Dirac delta function at $\omega=0$. Including the Kramers-Kronig transformation of this contribution leads to
\begin{equation}\label{sigma_2}
\sigma_2(\omega,T) = KK[\sigma_1](\omega,T)+\frac{W_{sc}(T)}{\pi\omega},
\end{equation}
where $KK[\sigma_1]$ is the Kramers-Kronig transformation of Eq.~\eqref{sigma_1} and $W_{sc}$ is the optical weight transferred into the delta function at zero frequency. $W_{sc}$ can be obtained from the sum rule
\begin{equation}
\frac{W_{sc}}{2}=\int_{0^+}^\infty(\sigma_1^N-\sigma_1^S)d\omega,
\end{equation}
where $\sigma_1^N$ and $\sigma_1^S$ are the optical conductivities in the normal and superconducting states, respectively. This standard rule from BCS theory has been shown to hold in the YRZ model as well \cite{carbotte:2010}.

As we can see in Eq.~\eqref{sigma_1}, the conductivity depends on the product of two spectral functions, $A(\k,\omega')$ and $A(\k,\omega'+\omega)$. In the superconducting state, there is also the similar product of Gorkov anomalous functions $B(\k,\omega')$ and $B(\k,\omega'+\omega)$. These products describe the transition from energy $\omega'$ to $\omega'$ plus the photon energy $\omega$. Each spectral function describes two branches, $\alpha=+$ or $-$, with respective weights $W_\k^+$ and $W_\k^-$. The product of two spectral functions hence contains four terms, which we will denote by $++$, $--$, $+-$, and $-+$. The first two of these, $++$ and $--$, describe intraband transitions within the $+$ or $-$ band; the latter two describe interband transitions.

We can easily determine that the intraband transitions create a Drude response centered at $\omega=0$, while the interband transitions create a peak centered at some finite frequency. Evaluating Eq.~\eqref{sigma_1} in the clean limit ($\Gamma\to0$), we find $\sigma_1=\sigma_D+\sigma_{IB}$, where
\begin{align}
\sigma_D &= -2\pi e^2g_t^2\sum_\k\sum_{\alpha=\pm} v^2_{k_x}(W_\k^\alpha)^2\frac{\partial f(E_S^\alpha)}{\partial E_S^\alpha}\delta(\omega),\label{Drude}\\
\sigma_{IB} &= 2\pi e^2g_t^2\sum_\k  v^2_{k_x}W_\k^+W_\k^-\nonumber\\
&\quad \times\Bigg\lbrace(u^-u^+-u^+v^-)^2\frac{1-f(E_S^+)-f(E_S^-)}{E_S^++E_S^-}\nonumber\\
&\quad \times\left[\delta(\omega-E_S^+-E_S^-)+\delta(\omega+E_S^++E_S^-)\right]\nonumber\\
&\quad -(u^+u^-+v^+v^-)^2\frac{f(E_S^+)-f(E_S^-)}{E_S^+-E_S^-}\nonumber\\
&\quad \times\left[\delta(\omega-E_S^++E_S^-)+\delta(\omega+E_S^+-E_S^-)\right]\Bigg\rbrace.\label{interband}
\end{align}
The term $\sigma_D$, peaked at $\omega=0$, is the Drude response coming from intraband transitions, while the term $\sigma_{IB}$, with contributions at finite energies $\pm E_S^+\pm E_S^-$, is the interband contribution arising from transitions between $\pm E_S^+$ and $\pm E_S^-$. In the non-superconducting case, we can reproduce the exact numerical results with impurities (i.e. those produced with Lorentzians of half-width $\Gamma/2$ in the spectral functions) by broadening the delta functions in the clean-limit formulas into Lorentzians of half-width $\Gamma$. (One should note that in the superconducting case at low temperature, this procedure fails, since there is a discontinuity at $\Gamma=0$. The difference between the superconducting and normal states can be understood from the fact that at $T=0$, $\frac{\partial f(E_S^\alpha)}{\partial E_S^\alpha}=-\delta(E_S^\alpha)$; this delta function, which arises only in the limit $\Gamma\to0$, has support on a surface if $\Delta_{sc}=0$ but only at the nodal point if $\Delta_{sc}\neq0$.)

\begin{figure}[tb]
\includegraphics{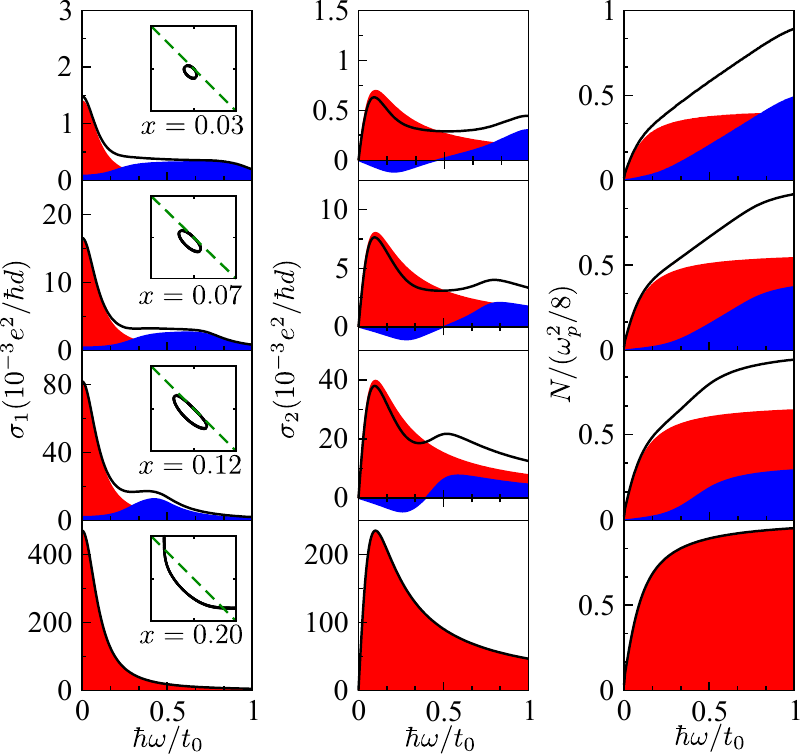}
\caption{(Color online) The optical conductivity at various dopings in the non-superconducting pseudogap state at temperature $T=0.01t_0/k_B$ and with impurity scattering $\Gamma=0.1t_0$. Left column: the real part $\sigma_1(\omega)$ of the conductivity as a function of frequency. The insets show the Fermi surface in the upper-right quadrant of the first Brillouin zone; the dashed green line indicates the boundary of the antiferromagnetic Brillouin zone. Middle column: the corresponding imaginary part $\sigma_2(\omega)$. Right column: the partial optical sum $N(\omega)$, normalized to one-eighth the square of the plasma frequency. In each plot, the red shaded region indicates the contribution from intraband scattering; the blue, from interband.}
\label{basov}
\end{figure}

\begin{figure}[tb]
\begin{center}
\includegraphics{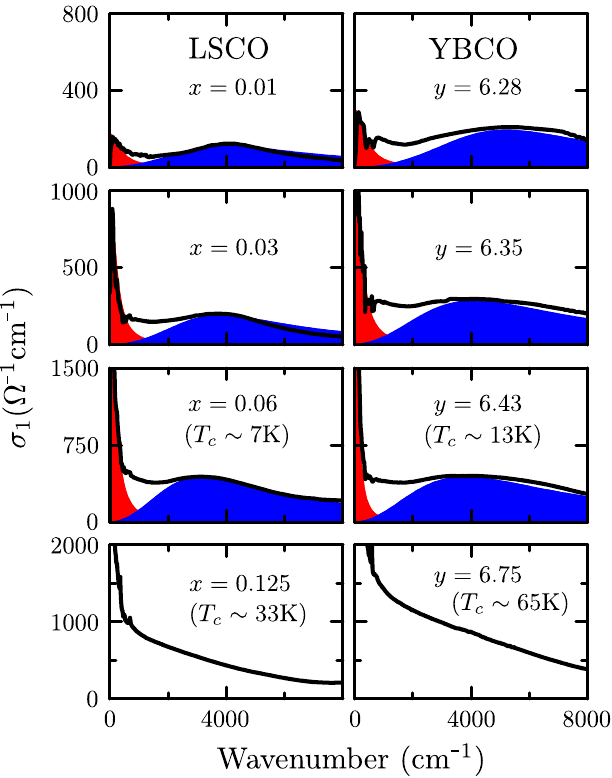}
\end{center}
\caption{(Color online) Experimental results for optical conductivity in LSCO (left panels) and YBCO (right panels) with a sequence of dopings. Red shaded regions indicate Drude contributions, while blue regions indicate contributions from mid-infrared Lorentzian oscillators. The top two sets of panels are for nonsuperconducting crystals at 10K. The other panels are for superconducting crystals at $T\simeq T_c$ (or $T^*$, the temperature at which the pseudogap opens, for $y=6.75$). This figure is taken from Lee et al.~\cite{lee:2005} with permission from the authors. \label{Basovs_fig}}
\end{figure}

In the left-hand column of Fig.~\ref{basov}, we display numerical results for $\sigma_1$ as a function of frequency in the non-superconducting state. (Because we have restricted ourselves to a CuO$_2$ plane, the conductivity takes on units $\frac{e^2}{\hbar d}$, where $d$ is the distance between CuO$_2$ planes.) We focus on the behavior as doping is decreased. At $x=0.20$ there is no pseudogap, and the Fermi surface (shown in the inset) is the familiar surface of Fermi liquid theory. In this case there is only one band, so only intraband transitions contribute to the conductivity. And the conductivity profile, shown in black, closely follows the well-known Drude form, with only slight modifications due to our model's more complicated band structure, which is restricted to the CuO$_2$ Brillouin zone rather than the usual continuum approximation with infinite bands. At dopings smaller than $x=0.20$, there is a finite pseudogap, and the Fermi surface is reconstructed into contours defining Luttinger hole pockets (again shown in the insets). In these cases the two bands $\alpha=\pm$ appear, leading to both an intraband contribution (indicated by the red shaded regions) and an interband one (indicated by the blue shaded regions).

Analogous sequences of results for a range of dopings were obtained experimentally by W.J. Padilla et al. \cite{padilla:2005} and Y.S. Lee et al. \cite{lee:2005}. We show the experimental data for LSCO and YBCO in Fig.~\ref{Basovs_fig}, reproduced from Ref.~\cite{lee:2005}. In that figure, the blue shaded regions indicate all sources of mid-infrared absorption, including incoherent scattering and possibly charge- or spin-density waves, not only the interband contribution of our model. However, the red shaded regions unambiguously show the Drude conductivity, the magnitude of which is a measure of the remaining coherence in the system as the doping is decreased toward the Mott insulating state at $x=0$. Our results are in agreement with the data, showing the Drude response greatly diminishing with decreasing $x$ (note the changing scale on the $y$-axis in both figures), but remaining finite so long as $x$ is finite. One can easily understand this within our model: As $x$ is decreased, the overall scale of the conductivity is greatly decreased, primarily due to the Gutzwiller factor $g_t^2$ appearing in Eqs.~\eqref{Drude} and \eqref{interband} but also due to the decreasing size of the Luttinger hole pocket, which reduces the size of the Fermi surface and the number of available charge carriers. But the hole pocket remains of finite size for $x>0$, which allows the coherent response to persist. 

In the middle column of Fig.~\ref{basov}, we display the corresponding imaginary part of the conductivity. In the pseudogap regime there are two distinct contributions to this quantity as well. The intraband contribution is as expected for a Drude response, but the interband piece is quite different in form, becoming negative at small $\omega$.

In the right-hand column, we display the partial optical sum
\begin{equation}
N(\omega)=\int_0^\omega\sigma_1(\omega')d\omega'.
\end{equation}
We have normalized this quantity to the total spectral weight $\omega_p^2/8=\int_0^\infty\sigma_1(\omega')d\omega'$, where $\omega_p$ is the plasma frequency. For the Fermi liquid case ($x=0.2$), a single energy scale is present, while in the pseudogap state a second scale arises, indicated by the small humps in the curves. From the blue and red shaded regions, we see that the interband transitions contribute a larger fraction of the optical weight as the doping is decreased; eventually, as shown in the upper-right-hand frame, the interband contribution dominates over the intraband one. Nevertheless, in the YRZ model, the Drude response persists for all positive values of doping.

We can determine more about the approach to the insulating state by examining two additional quantities: the effective optical mass $\m$ of the charge carriers, and the effective number $\n$ of them. The effective optical mass is defined by a generalized Drude form, which has become fairly standard in discussions of the optical conductivity of highly correlated systems. We write the conductivity in terms of $\m$ and an optical scattering rate $1/\tau$:
\begin{equation}
\sigma(\omega)=\frac{\omega_p^2}{4\pi}\frac{1}{1/\tau(\omega)-i\omega\m(\omega)/m},
\end{equation}
where $m$ is the bare band mass. From this formula, we see that $\m$ is given by
\begin{equation}\label{m_eff}
\frac{\m(\omega)}{m}=\frac{\omega_p^2}{4\pi}\frac{1}{\omega}\frac{\sigma_2}{\sigma_1^2+\sigma_2^2}.
\end{equation}
Here $\m$ is a function of frequency, but we will be interested only in the limit $\lim_{\omega\to0}\m(\omega)/m$; henceforth, $\m/m$ will denote this limiting value.

\begin{figure}[tb]
\includegraphics{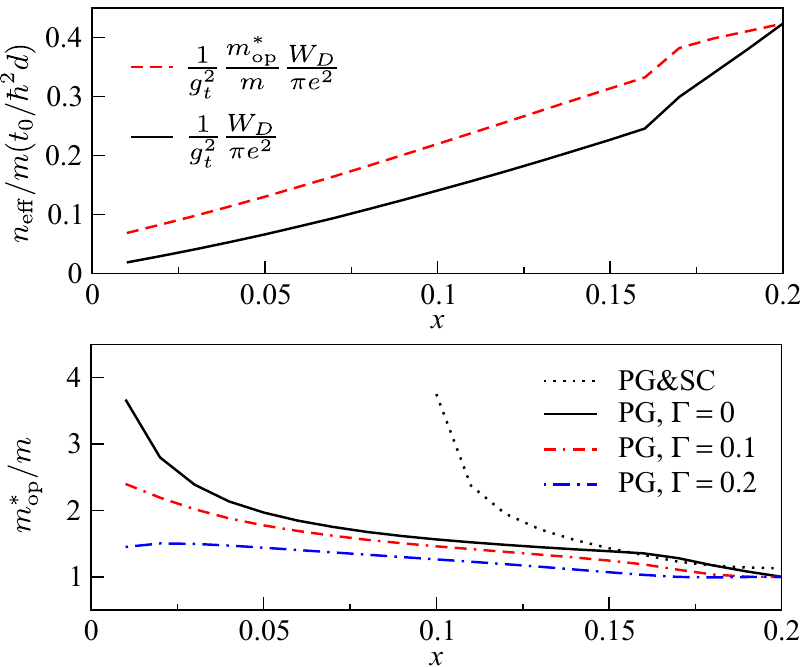}
\caption{(Color online) Top: the effective number $\n$ of charge carriers as a function of doping $x$ in the non-supercondicting state with no impurity scattering. For the dashed red curve, the renormalized optical mass is used to define $\n$; for the solid black curve, the bare mass is used. Bottom: the optical effective mass $\m$ as a function of doping, normalized to the bare mass $m$. Note the dependence on the residual impurity scattering rate $\Gamma$ (expressed here in units of $t_0$). The dotted curve is in the superconducting state in the clean limit, and it diverges at the end of the superconducting dome, corresponding to $x\approx 0.09$ in our model's phase diagram. All curves in both panels are for zero temperature.}
\label{meff_and_neff}
\end{figure}

Our numerical results for $\m$ as a function of doping are shown in the lower frame of Fig.~\ref{meff_and_neff}. The mass depends on frequency, temperature, impurity scattering, and the state of the system (i.e. whether or not a pseudogap is present, and whether or not the state is superconducting). When the system is non-superconducting, the effective mass becomes equal to the bare mass at $x=0.2$, where the pseudogap vanishes. When a pseudogap is present at lower dopings, the effective mass increases, but it changes by only a factor of 1--4 across the phase diagram. Furthermore, this increase is suppressed by impurity scattering: when $\Gamma=0.1t_0$ and $\Gamma=0.2t_0$, the effective mass is roughly constant, and clearly will not diverge in the approach to $x=0$. (Temperature causes an analogous suppression, but only at the experimentally irrelevant values $T\gtrsim0.1t_0/k_B$.)

Although our primary interest in $\m$ is its behavior in the non-superconducting state, we also present results when both gaps are present (dotted curve in Fig.~\ref{meff_and_neff}). In this case the mass diverges at the bottom of the superconducting dome, at $x\approx0.9$ in our phase diagram. Also, the effective mass does not equal the bare mass at $x=0.2$. However, when the system is superconducting, the effective mass defined by \eqref{m_eff} no longer has a straightforward interpretation as an electron mass. Instead, it is related to the superfluid density. Specifically, it is the ratio of total optical weight to the optical weight contributed by the superfluid:
\begin{equation}\label{mass_SC}
\frac{\m}{m}=\frac{\omega_p^2/4}{W_{sc}},
\end{equation}
which follows immediately from Eq.~\eqref{m_eff}, given that the term $\frac{W_{sc}}{\pi\omega}$ in Eq.~\eqref{sigma_2} blows up in the limit $\omega\to0$. At the bottom of the superconducting dome, the superfluid density vanishes, causing the right-hand side of Eq.~\eqref{mass_SC} to diverge.

We now define an effective number of charge carriers involved in the Drude response, beginning with the standard formula $n=\frac{m\omega_p^2}{4\pi e^2}$. Because we are interested only in the carriers involved in the Drude response, we replace $\omega_p^2$ with $4W_D$, where $W_D$ is the Drude optical weight defined by
\begin{equation}
\frac{W_D}{2} = \int^\infty_0\sigma_D(\omega)d\omega.
\end{equation}
The conductivity contains a factor $g^2_t(x)$, where $g_t(x)$ gives the quasiparticle weight remaining in the electronic Green's function, with the rest transferred by correlations to an incoherent background. As we wish to count the effective number of carriers, we divide this factor out of our definition of $\n$. Furthermore, we wish to account for the changes of the effective mass away from the band mass, so we use $\m$ in place of $m$. Our definition is thus
\begin{equation}
\n = \frac{1}{g_t^2}\frac{\m W_D}{\pi e^2}.
\end{equation}
The dashed red curve in the upper frame of Figure \ref{meff_and_neff} shows $\n$ as a function of doping. For comparison, we also show (in solid black) the result when the band mass $m$, rather than $\m$, is used in the definition of $\n$. In both cases, the defined number of charge carriers decreases toward zero in the approach to the insulating state. (Though the dashed curve appears to be approaching a finite value, this is a misleading artefact of the relatively large increase in effective mass in the clean limit; since the two curves differ by a factor of the effective mass, so long as the effective mass does not diverge, the dashed curve must approach zero if the solid curve does.) Furthermore, below $x\approx0.16$, where only the hole pockets exist, the effective number of charge carriers is directly proportional to the doping, confirming that our definition of $\n$ is appropriate. Above $x\approx0.16$, the behavior changes as the electron pockets appear together with the hole pockets.

We can now clearly understand the essential features of the optical response in deeply underdoped cuprates. The Drude conductivity decreases rapidly as $x\to0$, but it remains finite for all positive values of $x$. In the foregoing discussion, we have identified three factors governing this behavior. First, the square of the Gutzwiller factor $g_t(x)$ tells us that the remaining weight in the quasiparticle pole decreases as correlations grow and transfer spectral weight to the incoherent background. Second, the number of charge carriers drops linearly with decreasing doping. Third, the optical effective mass increases. However, only the first two factors have a large impact on the conductivity; the increased mass has only a modest impact, since the carriers become heavier by only a factor of roughly 1--4 at dopings as low as 1\%.

\section{DC conductivity, Wiedemann-Franz law, and microwave conductivity}\label{DC conductivity}
In the preceding section we focused on the non-super\-conducting state. When a superconducting gap is included, all points on the Luttinger pocket contours become gapped except those in the nodal direction. But from Eq.~\eqref{Wpm} we see that the weight $W_\k^-$ associated with the backside of the pocket (i.e. the side facing the antiferromagnetic Brillouin zone boundary) is zero in the nodal direction. Hence, there is effectively only one point that remains ungapped: the point in the nodal direction on the side of the pocket facing toward the center of the Brillouin zone. This is referred to as the Dirac point. At very low temperatures and photon frequencies, this is the only active point. But this point and those in its very near vicinity are largely unaffected by pseudogap formation, since the pseudogap vanishes in the nodal direction. Therefore, even though the pseudogap radically reconstructs the Fermi surface into small pockets, at sufficiently low temperatures it has negligible effect on low-frequency transport properties such as the DC electrical conductivity, thermal conductivity, and microwave conductivity. In this section we show the validity of these statements and the effect of the pseudogap as the temperature is increased.

\begin{figure}[tb]
\includegraphics[clip=true]{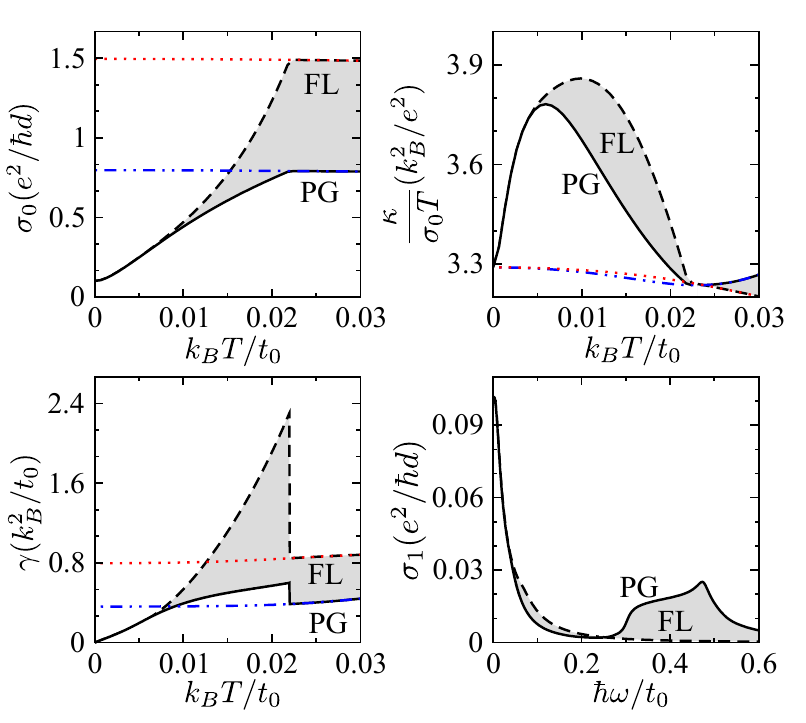}
\caption{(Color online) Conductivity and specific heat in the superconducting state, with a pseudogap (solid curves) and without (dashed curves). Dotted red and dot-dashed blue curves are in the non-superconducting state. Shaded regions emphasize the difference between the pseudogap and Fermi liquid results. Top left: the DC electrical conductivity $\sigma_0\equiv\sigma_1(\omega=0)$ as a function of temperature. Top right: the corresponding ratio of thermal conductivity $\kappa/T$ to the DC conductivity. Lower left: the specific heat $\gamma$ as a function of temperature. Lower right: the real part of the optical conductivity as a function of frequency. All curves are for $x=0.12$ and $\Gamma=0.01t_0$}
\label{DCtransport}
\end{figure}

We obtain the DC conductivity by letting $\omega\to0$ in \eqref{sigma_1}, yielding
\begin{align}
\sigma_0(T) &= -2\pi e^2\sum_\k v^2_{k_x}\int_{-\infty}^\infty d\omega'\frac{\partial f(\omega')}{\partial\omega'}\nonumber\\
&\quad\times\left[A(\k,\omega')^2+B(\k,\omega')^2\right],
\end{align}
where $\sigma_0\equiv\lim_{\omega\to0}\sigma_1(\omega)$. The thermal conductivity is given by the similar formula
\begin{align}
\frac{\kappa(T)}{T} &= -2\pi \frac{k_B^2}{T^2}\sum_\k v^2_{k_x}\int_{-\infty}^\infty d\omega'\frac{\partial f(\omega')}{\partial\omega'}\omega'^2\nonumber\\
&\quad\times\left[A(\k,\omega')^2-B(\k,\omega')^2\right].
\end{align}
The top-left frame of Fig.~\ref{DCtransport} shows $\sigma_0$ as a function of $T$. We consider two cases: the results of the YRZ model with a pseudogap (solid curve) and the results with $\Delta_{pg}$ set to zero (dashed curve). As we expect, at low temperatures the two curves are indistinguishable, since the pseudogap has not changed the environment about the Dirac point. As the temperature increases, however, the DC conductivity in the superconducting state is suppressed below its Fermi liquid value, since the pseudogap gaps out states that would otherwise be available. The shaded region in the figure emphasizes this effect. Since the thermal conductivity exhibits the same behavior, we do not explicitly show it. Instead, in the upper-right frame, we show the Wiedemann-Franz ratio $\frac{\kappa}{\sigma_0T}$. Again the shaded region emphasizes the difference between results with and without a pseudogap, and again, at low temperatures the pseudogap is seen to have no effect. Analogous results were found by Borne et al. \cite{borne:2010} in their study of specific heat $\gamma\equiv c_V/T$, and for comparison we include those results in the lower-left frame.

As mentioned previously, we expect the same behavior in the case of the microwave conductivity: given a low temperature, if the photon frequency is sufficiently small, the pseudogap should have a negligible effect. Our expectations are confirmed by a plot of $\sigma_1$ as a function of frequency at zero temperature, shown in the lower-right frame of Fig.~\ref{DCtransport}. As we expect, the curve in the Fermi liquid case (dashed) merges with the curve in the pseudogap case (solid) at low frequencies. For the case shown, the pseudogap becomes important only in the frequency range where interband transitions predominate; for the Fermi liquid case, in which no such transitions occur, the conductivity behaves simply as a Drude tail in this region.

\begin{figure}[tb]
\includegraphics{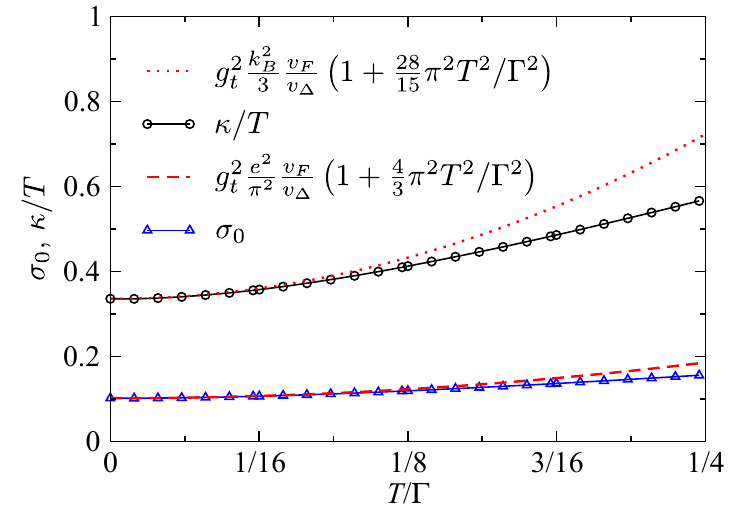}
\caption{(Color online) DC electrical conductivity $\sigma_0$ (open blue triangles) and thermal conductivity $\kappa/T$ (open black circles) as a function of temperature (in units of residual scattering rate $\Gamma$) in the superconducting state. The dashed and dotted red lines are analytic forms valid in the limit $T/\Gamma\to0$. All curves are for $x=0.12$ and $\Gamma=0.005t_0$.}
\label{lowT}
\end{figure}

Figure \ref{lowT} displays results for the very-low-temperature behavior of electrical and thermal conductivity. In a previous paper, Carbotte \cite{carbotte:submitted} derived an analytic expression for the universal (i.e. impurity-independent) zero-temperature limits of both thermal and electrical conductivity in a $d$-wave superconductor with a pseudogap. These limits are important signatures of nodes in the superconducting order parameter \cite{taillefer:1997,sutherland:2003}. Including the leading-order temperature correction (which shows the first dependence on impurity content), the limiting behavior is given by
\begin{align}
\sigma_0 &= g_t^2\frac{e^2}{\pi^2}\frac{v_F}{v_\Delta}\left(1+\frac{4}{3}\pi^2\frac{T^2}{\Gamma^2}+\ldots\right),\label{sigma_limit}\\
\frac{\kappa}{T} &= g_t^2\frac{k_B^2}{3}\frac{v_F}{v_\Delta}\left(1+\frac{28}{15}\pi^2\frac{T^2}{\Gamma^2}+\ldots\right),\label{kappa_limit}
\end{align} 
where $v_F=\big|\frac{\partial\xi_\k}{\partial\k}\big|_{\mathcal{P}_D}$ and $v_\Delta=\big|\frac{\partial\Delta_{sc}}{\partial\k}\big|_{\mathcal{P}_D}$ are respectively the Fermi and gap velocities at the Dirac point $\mathcal{P}_D$, and we have set $\hbar=d=k_B=1$, except in the overall prefactor of $k_B^2$. Except for the important factor of $g_t^2$, these formulas are unaltered from standard results for $d$-wave superconductors \cite{lee:1993,graf:1996,durst:2000}. They are derived by performing a Sommerfeld expansion and expanding around $\mathcal{P}_D$. In that derivation, one must assume the residual impurity scattering rate $\Gamma$ is much smaller than all energy scales other than $T$; for small scattering rates, one can show that Eqs.~\eqref{sigma_limit} and \eqref{kappa_limit} are accurate to within an error of order $\Gamma^2$. We see in Fig.~\ref{lowT} that the analytic expressions yield excellent agreement with numerical evaluations of the exact formulas. As we have shown above, because the pseudogap shares the $d$-wave symmetry of the superconducting gap, these results are valid regardless of whether or not the pseudogap is present. Note that these results hold only for finite $\Gamma$; as mentioned in the preceding section, in the superconducting state there is a discontinuity at $(\Gamma=0,T=0)$, and if $\Gamma$ vanishes identically, the conductivity at $T=0$ also vanishes identically, rather than reaching the universal value given above.

All derivations of these limits involve only the coherent part of the electron propagator. However, Schachinger and Carbotte \cite{schachinger:1998} considered the case of coupling to a boson spectrum within Eliashberg theory. Such a calculation fully accounts for the inelastic scattering that provides the boson-assisted incoherent contribution to the conductivity. They showed analytically that in this case Eqs.~\eqref{sigma_limit} and \eqref{kappa_limit} still hold, but with the factor of $g_t^2$ replaced by $1/(1+\lambda)$, where $\lambda$ is the mass-enhancement factor associated with the electron-boson interaction. Like $g_t^2$, the factor $1/(1+\lambda)$ measures the reduction of the coherent part of the conductivity due to interactions.

\begin{figure}[tb]
\includegraphics{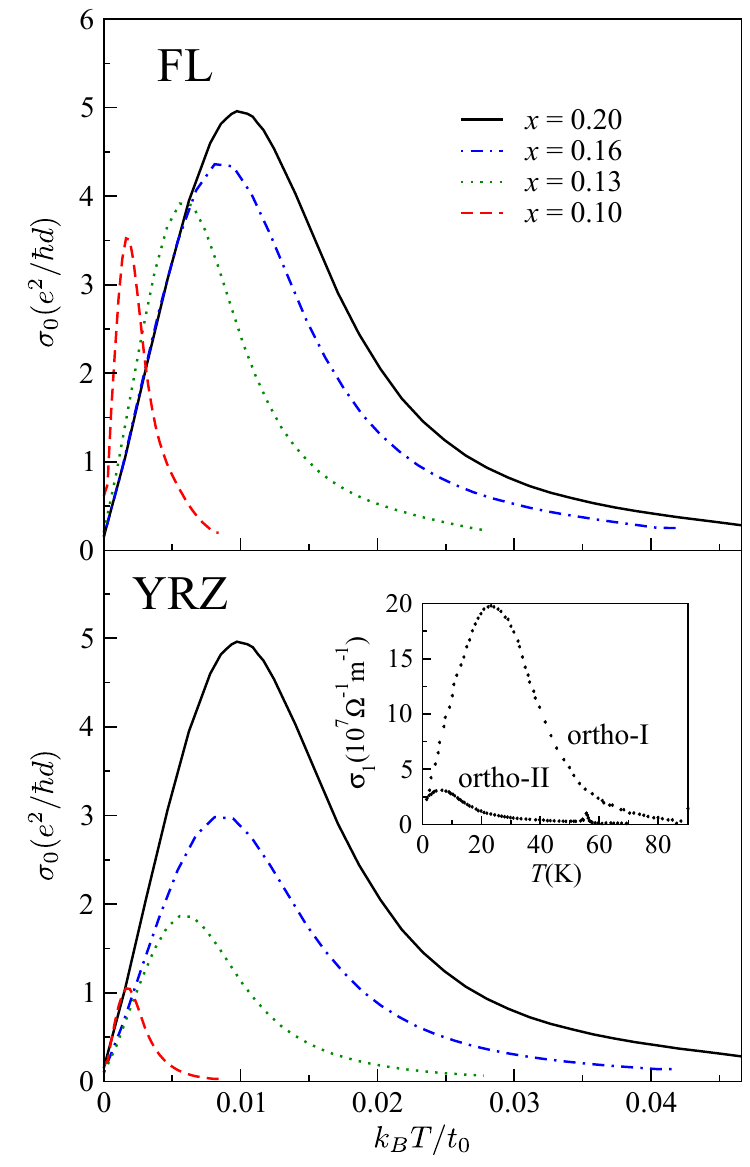}
\caption{(Color online) DC conductivity $\sigma_0$ in the superconducting state as a function of temperature for four values of doping $x$, with a temperature dependent scattering rate $\Gamma=0.001t_0+0.16t_0(T/T_c)^4$.  Top: results in the Fermi liquid model. Bottom: results in the YRZ model. Results in the Fermi liquid model are related to those in the YRZ model by setting the pseudogap to zero and dropping all Gutzwiller coherence factors (except those that define the band structure). The inset in the bottom frame shows experimental microwave conductivity results for ortho-I and -II YBCO, taken from R. Harris et al. \cite{harris:2006}. For ortho-I, two samples were considered, with critical temperatures $T_c=89$K and $91$K. For ortho-II, $T_c=56$K. All experimental data is for $a$-axis conductivity at 1.1GHz.}
\label{microwave}
\end{figure}

We now move to our final results, shown in Fig.~\ref{microwave}. These are again plots of the DC conductivity as a function of temperature, but unlike in the preceding plots, which used a constant impurity scattering rate, here we have used a temperature-dependent rate in order to account for inelastic scattering. This allows us to meaningfully compare with experimental data on microwave conductivity: it is now widely accepted that the large peak, usually at a temperature much below $T_c$, in the microwave conductivity \cite{nuss:1991,bonn:1992,bonn:1994,hosseini:1999} and thermal conductivity \cite{yu:1992,matsukawa:1996} of cuprates is not a coherence peak associated with superconductivity, but rather the effect of a rapid decrease in inelastic scattering with decreasing temperature below $T_c$. One prominent model that has emerged from fits to data is a $T^4$ power law \cite{hosseini:1999,marsiglio:2006}, and here we have used
\begin{equation}
\Gamma(T)=0.001t_0+0.16t_0(T/T_c)^4
\end{equation}
for all values of doping. In the upper frame of Fig.~\ref{microwave}, we show results for Fermi liquid theory, with no pseudogap present and no overall factor of $g_t^2$ in the conductivity. (However, the Gutzwiller factors defining the band structure are left intact, and we have scaled the vertical axis by the Gutzwiller factor at optimal doping to allow easy comparison with the YRZ model.) In the lower frame, we show results for the YRZ model. In both cases, we show a range of dopings, and we see that though the shapes of the curves are very similar in both models, the Gutzwiller factor in the YRZ model causes the magnitude of the conductivity to decrease much more rapidly with decreasing doping. (We have confirmed numerically that the pseudogap has only a negligible effect in comparison.)

The inset in the lower frame shows experimental data for microwave (1.1GHz) conductivity, taken from Harris et al. \cite{harris:2006}. The upper data set was obtained from two samples of optimally doped ortho-I YBa$_2$Cu$_3$O$_{6.993}$ with $T_c$ values of $89$K and $91$K; the lower set, from a sample of ortho-II YBa$_2$Cu$_3$O$_{6.5}$ with $T_c=56$K. For ortho-II, every second chain is empty. Given our choice of $x=0.2$ as optimal doping, ortho-II would correspond to a doping between $x=0.10$ and $x=0.13$. Comparing these data sets with the corresponding numerical curves, we see that the YRZ model, with its large decreases in magnitude with decreasing doping, yields far better qualitative agreement with experiment than does the Fermi liquid model. In order to match the experimental data, we could alter the parameters in the scattering rate, and in particular, we could allow them to differ between ortho-I and ortho-II. (We could also generate numerical results at 1.1GHz, rather than DC data, which would slightly lower our curves.) However, here we are interested only in qualitative behavior, and our results clearly show that the data is more easily understood within the YRZ model than ordinary Fermi liquid theory. Furthermore, we see that the Gutzwiller factor is the essential element in explaining the data.

\section{Summary and conclusions}\label{summary}
We have found that the approach to the insulating Mott state captured in the resonating valence bond spin liquid model developed by Yang, Rice, and Zhang is in qualitative agreement with the experimental observation that a coherent Drude response persists in the deeply underdoped regime of the cuprates down to the 1\% doping level. In this regime, small hole Luttinger pockets remain, and on the boundary contours of these pockets zero energy excitations exist, leading to a Drude term in the conductivity. This term decreases as the size of the pockets (and therefore the number of charge carriers) shrinks with decreasing doping. It also decreases due to the presence of Gutzwiller factors accounting for increased correlations, which cause a transfer of spectral weight to an incoherent background. An increase in optical effective mass also causes a decrease in metallicity as the insulating state is approached, but in agreement with experiment, this is not a dominant effect: in the regime of interest, the effective mass is only of order twice the band mass. In this sense, the charge carriers do not become heavy.

We have also investigated the effect of pseudogap formation on the temperature dependence of the DC and thermal conductivity in the superconducting state. If the Gutzwiller factor is not considered, the pseudogap leaves unaltered the low-temperature  behavior of the conductivities, even though it causes a radical reconstruction of the Fermi surface, from the large open contour of Fermi liquid theory to small Luttinger pockets. This observation is explained by the fact that the zero-temperature conductivities depend only on the region of reciprocal space around the tip of the Dirac cone, which is unaffected by the finite pseudogap because the pseudogap has $d$-wave symmetry, vanishing along the nodal direction where the Dirac point lies. Due to this, the universal limit associated with $d$-wave superconductivity survives in the pseudogap regime. However, the universal limit is now modified by a Gutzwiller factor $g^2_t(x)$. The same holds true for the approach to zero temperature: analytic formulas derived for the small-$T$ limit are unaltered except for the appearance of a Gutzwiller factor.

Finally, we have considered microwave conductivity. Plots of conductivity as a function of temperature show that Fermi liquid theory and the YRZ model yield curves of very similar shape but greatly different magnitude. Comparing these numerical results to existing data for ortho-I and -II YBCO, we found that only the YRZ model is in qualitative agreement with experiment. The lower conductivity of ortho-II as compared to ortho-I is seen to result primarily from the increased correlations described by the Gutzwiller factor.

\begin{acknowledgement}
We wish to thank D.~N. Basov and Y.~S. Lee for allowing us use of a figure from Ref.~\cite{lee:2005}. This work was supported in part by the Natural Sciences and Engineering Research Council of Canada and the Canadian Institute for Advanced Research.
\end{acknowledgement}

\bibliography{optic_YRZ_epjb}

\end{document}